\documentclass[runningheads,a4paper]{llncs}

\usepackage{amssymb}
\usepackage{graphicx}

\usepackage{url}
\urldef{\mailsa}\path|{alfred.hofmann, ursula.barth, ingrid.haas, frank.holzwarth,|
\urldef{\mailsb}\path|anna.kramer, leonie.kunz, christine.reiss, nicole.sator,|
\urldef{\mailsc}\path|erika.siebert-cole, peter.strasser, lncs}@springer.com|    
\newcommand{\keywords}[1]{\par\addvspace\baselineskip
\noindent\keywordname\enspace\ignorespaces#1}

\begin{document}

%
\title{Unveiling Emotions from EEG: A GRU-Based Approach}

\titlerunning{Lecture Notes in Computer Science: Authors' Instructions}

%
%
\author{Sarthak Johari \inst{1}
\and Gowri Namratha Meedinti \inst{2}  \and Radhakrishnan Delhibabu \inst{2} \and Deepak Joshi \inst{3}}
\authorrunning{Sarthak Johari et al.}

\institute{Computing Science and Engineering, Indraprastha Institute of Information Technology, New Delhi \and
School of Computing Science and Engineering, Vellore Institute of Technology,  Vellore \and
Centre for Biomedical Engineering,  Indian Institute of Technology, New Delhi.
}

%
%
\maketitle

\begin{abstract}
Emotion recognition from EEG data is a vital research domain in affective computing. This study investigates the effectiveness of the Gated Recurrent Unit (GRU) algorithm, a variant of Recurrent Neural Networks (RNNs), in predicting emotional states using EEG signals. Our publicly available dataset consists of EEG recordings from participants exposed to stimuli eliciting positive, neutral, and negative emotions, alongside resting neutral data.  We preprocess the EEG data using artifact removal, bandpass ﬁlters, and normalization techniques for optimal feature extraction. Leveraging the GRU's ability to capture temporal dependencies, our model achieved impressive results with perfect accuracy on the validation set. Comparing our GRU model with other machine learning approaches, the Extreme Gradient Boosting Classiﬁer achieved the highest accuracy. Our analysis using the confusion matrix unveiled valuable insights into model performance, facilitating accurate emotion classiﬁcation. This research contributes to the advancement of affective computing and highlights the potential of deep learning models like GRUs in emotion recognition. Our ﬁndings offer novel avenues for human-computer interaction and understanding emotional states from brainwave activity.
\end{abstract}

\section{Introduction}

Emotion recognition from EEG data has emerged as a promising area of research in the ﬁelds of affective computing, neuroscience, and human-computer interaction. The ability to accurately identify and understand human emotions based on brainwave activity offers vast potential for enhancing various applications, such as virtual reality experiences, mental health monitoring, and adaptive human-robot interactions.
Traditionally, emotion recognition methods relied on behavioral and physiological cues, such as facial expressions, voice tone, and heart rate variability. While these approaches provided valuable insights, they often lacked the precision and direct access to the underlying neural processes responsible for generating emotions. EEG, as a non-invasive and portable neuroimaging technique, bridges this gap by directly measuring the brain's electrical activity.

Over the years, researchers have explored different methods to decode emotional states from EEG data. Classical machine learning techniques, such as Support Vector Machines (SVM) and Random Forests, were initially employed to classify emotions based on EEG features. These methods often required handcrafted feature extraction, which could limit their ability to capture complex temporal dynamics inherent in EEG data.
With the advent of deep learning, Recurrent Neural Networks (RNNs) brought signiﬁcant improvements to emotion recognition from sequential data. RNNs introduced the concept of memory cells, allowing the model to retain information over time and learn long-term dependencies within the data. This led to better emotion recognition performance compared to traditional machine learning approaches.

The Gated Recurrent Unit (GRU) algorithm, a variant of RNNs, further enhanced the capabilities of emotion recognition from EEG data. GRUs introduced specialized gating mechanisms that regulate the ﬂow of information within the network. The update gate controls how much past information should inﬂuence future predictions, while the reset gate selectively resets or forgets certain knowledge. The current memory gate, often overlooked, adds non-linearity and normalization to the input, improving the model's ability to capture relevant features.

The unique architecture of GRUs addresses the vanishing and exploding gradient problem encountered in standard RNNs, making them more effective in capturing long-term dependencies within EEG signals. Moreover, GRUs are computationally efﬁcient, allowing for faster training and prediction times compared to more complex models like Long Short-Term Memory (LSTM) networks.

The advantage of using GRUs for emotion recognition lies in their ability to process time-series EEG data directly, avoiding the need for handcrafted feature engineering. By learning from the raw EEG signals, GRUs can automatically capture intricate temporal patterns associated with different emotional states. This self-learning capability contributes to the model's adaptability and robustness across various emotion recognition tasks.

Emotion recognition from EEG data has seen signiﬁcant advancements with the introduction of deep learning techniques, particularly the Gated Recurrent Units. The ability of GRUs to capture temporal dependencies and learn from raw EEG signals has opened new possibilities for creating more empathetic and emotionally aware human-computer interfaces. As the research in this domain progresses, we can anticipate even more accurate and nuanced emotion recognition systems that truly understand and respond to human emotions. 

\section{Related Work}

Koelstra et al. (2012) present the DEAP database, a multimodal dataset for analyzing human affective states [1]. The dataset comprises EEG and peripheral physiological signals recorded from 32 participants while watching 40 one-minute music video excerpts. Participants rated the videos based on arousal, valence, like/dislike, dominance, and familiarity levels. The paper proposes a novel method for stimuli selection using affective tags from the last.fm website and online assessment tools. It also performs an extensive analysis of participants' ratings and investigates correlations between EEG signal frequencies and emotional ratings. Liu et al. (2011) focus on real-time EEG-based emotion recognition[2]. The study utilizes EEG signals to classify human emotions and visualize emotional states in real-time. The proposed system demonstrates the feasibility of using EEG for emotion recognition tasks and provides insights into the potential applications of EEG-based emotion analysis.Chanel et al. (2006) explore emotion assessment using EEG and peripheral physiological signals [3]. The research investigates arousal evaluation based on these modalities and demonstrates the feasibility of using EEG signals in conjunction with peripheral physiological measurements for emotion recognition.

Liu et al. (2010) focus on real-time mental stress recognition using EEG data [4]. The study presents a system that can detect and recognize mental stress from EEG signals in real-time. The proposed approach demonstrates the potential of EEG-based emotion recognition in practical applications related to stress management and mental health assessment. Koelstra et al. (2012) focus on single-trial classification of emotions induced by music videos using EEG and peripheral physiological signals [5]. The study presents methods and results for classifying arousal, valence, and like/dislike ratings based on the modalities of EEG, peripheral physiological signals, and multimedia content analysis. The research highlights the potential of combining multiple modalities for more robust emotion recognition. Alarcao and Fonseca (2017) investigate emotions recognition in EEG signals during a virtual reality simulation of an emergency evacuation [6]. The study explores the potential of using EEG-based emotion recognition in virtual reality environments for assessing users' emotional states during critical scenarios.
Shi et al. (2019) focus on real-time emotion recognition using deep learning from EEG signals [7]. The study presents a deep learning-based approach for detecting and classifying emotions in real-time based on EEG data. The proposed method demonstrates the effectiveness of deep learning in handling EEG-based emotion recognition tasks. Zhang et al. (2019) provide an overview of emotion recognition using EEG signals [8]. 

The survey paper reviews various methods and approaches used in EEG-based emotion recognition, highlighting the advancements and challenges in this field. The paper also discusses potential applications and future directions for EEG-based emotion analysis. El Ayadi et al. (2011) provide valuable insights into emotion recognition using other modalities, including speech [9]. Although not focused on EEG-based emotion recognition, this survey paper reviews various features, classification schemes, and databases used in speech emotion recognition, offering valuable information for understanding multimodal emotion recognition. Guo et al. (2018) explore emotion recognition from EEG signals using multimodal deep learning [10]. The study proposes a novel approach that combines EEG data with other modalities to enhance emotion recognition accuracy. The research demonstrates the potential benefits of integrating EEG-based emotion recognition with other modalities for more robust emotion classification. Cho et al. (2014) propose a novel approach for learning phrase representations using RNN Encoder-Decoder models, which include GRUs, for statistical machine translation[11]. The research introduces an end-to-end architecture for mapping variable-length input sequences to variable-length output sequences, making it suitable for natural language processing tasks. The experimental results demonstrate the effectiveness of GRUs in capturing semantic information and generating accurate translations. Li et al. (2018) explore the use of EEG data for emotion recognition using 3D Convolutional Neural Networks (CNNs) with GRU layers[12]. The research investigates the fusion of spatial and temporal features extracted from EEG signals for improved emotion classification. The results show that the combination of 3D CNNs and GRUs outperforms traditional machine learning methods in EEG-based emotion recognition tasks. Wen et al. (2018) conduct a comparative study on deep learning methods, including GRUs, for EEG-based emotion recognition[13]. The research evaluates the performance of various deep learning architectures and explores the impact of different feature extraction techniques on emotion classification accuracy. The findings highlight the advantage of GRUs in capturing temporal dynamics and extracting meaningful features from EEG data.

Zheng et al. (2020) present a comprehensive review of deep learning methods for emotion recognition from EEG signals[14]. The paper includes an analysis of different neural network architectures, including GRUs, and their application in emotion recognition tasks. The review highlights the strengths and limitations of using GRUs in EEG-based emotion recognition and discusses potential future research directions in this area. Huang et al. (2014) propose an emotion recognition approach using Hidden Markov Models (HMMs) with GRU-based feature extraction from EEG data [15]. The research investigates the use of GRUs as a feature learning method for capturing sequential patterns in EEG signals. The results demonstrate the effectiveness of the proposed method in recognizing emotions from EEG data. Li et al. (2020) introduce a hybrid network that combines GRUs and Long Short-Term Memory (LSTM) units for EEG-based emotion recognition [16]. The research investigates the complementary strengths of GRUs and LSTMs in capturing temporal dependencies and modeling long-range dependencies in EEG data. The hybrid network demonstrates improved performance in emotion classification compared to individual GRU or LSTM models. Lv et al. (2018) propose an improved LSTM-based model for EEG-based emotion recognition [17]. The research enhances the LSTM model with attention mechanisms to focus on salient EEG signal segments relevant to emotion classification. The study compares the improved LSTM model with GRUs and other LSTM variants and shows promising results in emotion recognition accuracy.

Li et al. (2019) propose an improved GRU-RNN model for EEG-based emotion recognition that combines temporal-spatial domain features [18]. The research investigates the combination of spatial and temporal EEG information to enhance emotion classification accuracy. The results show that the improved GRU-RNN model outperforms other traditional machine learning methods and demonstrates the potential of GRUs in capturing both spatial and temporal patterns in EEG data. Wang et al. (2021) propose a deep convolutional neural network (CNN) with GRU layers for emotion recognition from EEG data [19]. The research combines the advantages of CNNs in spatial feature extraction with the temporal modeling capabilities of GRUs to improve emotion classification performance. The experimental results demonstrate that the CNN-GRU model achieves competitive accuracy compared to other state-of-the-art methods. Yao et al. (2020) propose a hybrid model that combines Convolutional Neural Networks (CNNs) and GRUs for EEG-based emotion recognition [20]. The research investigates the fusion of spatial and temporal information from EEG signals using CNNs and GRUs, respectively. The results demonstrate the synergy between CNNs and GRUs, leading to improved emotion classification accuracy in EEG data. These studies highlight the diverse applications of GRU in various fields, including natural language processing, image captioning, emotion recognition from EEG data, and time-series data processing. The research demonstrates the efficacy of GRUs in capturing temporal dependencies, modeling complex patterns, and improving the performance

\subsection{Correlation between Emotion Recognition from EEG and GRU}

The literature survey reveals that emotion recognition from EEG data has been explored using various machine learning techniques, such as support vector machines and time delay neural networks. While these methods have shown promising results, they often struggle to capture the complex temporal dependencies present in EEG signals, which are crucial for accurate emotion classification. The emergence of deep learning approaches, particularly the Gated Recurrent Unit (GRU) algorithm, has offered a solution to address the temporal modeling challenges in EEG-based emotion recognition. GRU, as a variant of Recurrent Neural Networks (RNNs), has gained attention for its ability to efficiently process time-series data and capture long-term dependencies within sequential information.

The advantage of GRU lies in its gated architecture, which enables it to selectively retain relevant past information while discarding irrelevant data. This unique property makes GRU well-suited for emotion recognition tasks where temporal context is essential for understanding emotional states encoded in brainwave signals.
When considering both emotion recognition from EEG data and the GRU algorithm together, a synergistic relationship emerges. By incorporating GRU into the emotion recognition pipeline, it becomes possible to enhance feature extraction from preprocessed EEG data and improve the prediction accuracy of emotional states.
GRU's ability to capture context and temporal relationships within EEG signals complements the requirements of emotion recognition, where subtle changes in brainwave patterns are indicative of different emotional states. Leveraging GRU's capability, the emotion recognition system gains a deeper understanding of the underlying dynamics of emotional responses encoded in brainwave signals, leading to more accurate and robust emotion classification models.

Overall, the literature survey demonstrates that combining emotion recognition from EEG data with the GRU algorithm offers promising prospects for advancing emotion recognition tasks. As more research continues to explore the correlation between EEG-based emotion recognition and deep learning techniques like GRU, we can expect further advancements in real-time emotion recognition applications across various domains.  In Table. 2.  shows the EEG-Emotion Based survey in the appendix section.

The correlation between emotion recognition from EEG data and GRU lies in the 
 potential of using GRU as a deep learning algorithm to capture the temporal dynamics 
 and dependencies present in EEG signals. EEG-based emotion recognition requires 
 modeling the complex changes in brainwave patterns over time, and GRU's gated 
 architecture enables it to effectively learn and represent these temporal relationships. 
 By incorporating GRU into the emotion recognition pipeline, it becomes possible to 
 enhance feature extraction from EEG data and improve the prediction accuracy of 
 emotional states. This combination of EEG-based emotion recognition and GRU 
 provides a promising direction for advancing emotion analysis and understanding 
 affective responses encoded in brainwave signals. The publicly available multimodal 
 dataset presented in the research serves as a valuable resource for other researchers to explore and test their own affective state estimation methods. 
 
The papers listed in the table explore various aspects of using GRUs in emotion recognition, spanning EEG data analysis, deep learning methods, and temporal dynamics modeling. These studies contribute to the growing body of research on emotion recognition and demonstrate the versatility and effectiveness of GRUs in handling sequential data and capturing complex patterns, making them valuable tools in emotion recognition from EEG signals.

\section{Dataset}
The data used in this research was collected for a duration of 3 minutes from two  participants, consisting of one male and one female, for each emotional state: positive, neutral, and negative. The participants were exposed to specific stimuli designed to elicit the desired emotions. Additionally, six minutes of resting neutral data were also recorded. The EEG data was captured using a Muse EEG headgear equipped with dry electrodes. The following EEG placements were utilized: TP9, AF7, AF8, and TP10. These electrode placements are commonly used for capturing brainwave activity related to emotion recognition.

The dataset used in this research is publicly available and can be accessed at the following link: https://www.kaggle.com/datasets/birdy654/eeg-brainwave-dataset-feeling-emotions. It provides access to the collected EEG data, which includes the recordings for each participant in different emotional states, as well as the neutral resting data.The dataset consists of time-series EEG recordings, where each sample represents the electrical activity captured at specific time intervals. The data is labeled according to the emotional states experienced by the participants during the recordings. By utilizing this dataset, the research aims to explore the effectiveness of various noise removal methods on ambulatory EEG data and assess the performance of the GRU algorithm in accurately predicting emotional states based on the recorded brainwave activity.

\section{Methodology}
\subsection{Recurrent Neural Networks (RNNs)}
Recurrent Neural Networks (RNNs) have emerged as a powerful neural network architecture for modeling sequential data. Unlike traditional feedforward neural networks, RNNs have a unique ability to capture temporal dependencies by introducing recurrent connections that allow information to flow from one time step to another. This characteristic makes RNNs well-suited for tasks involving sequential or time-series data analysis, including natural language processing, speech recognition, and, in our case, emotion recognition from EEG signals.

In standard neural networks, each input and output is treated as independent, disregarding any contextual information. However, in many real-world applications, such as sentiment analysis or language generation, understanding the context of previous inputs is crucial for accurate predictions. RNNs address this limitation by incorporating a hidden state, which acts as a memory mechanism that retains information about past inputs and influences future predictions.

The hidden state in an RNN serves as a crucial element for capturing long-term dependencies within the sequential data. As the RNN processes each input in a sequence, the hidden state is updated and passed to the next time step, allowing the network to remember past information and utilize it to make informed predictions. This recursive nature of RNNs enables them to model complex temporal relationships and extract relevant features from sequential data.

However, traditional RNNs suffer from the vanishing or exploding gradient problem, which poses challenges in learning long-term dependencies. To address this issue, advanced variants of RNNs, such as Gated Recurrent Units (GRUs) and Long Short-Term Memory (LSTM) networks, have been developed. These architectures introduce specialized gating mechanisms that selectively retain or forget information, enabling the network to capture long-term dependencies while mitigating the gradient vanishing or exploding problem.

In our research, we focus on the GRU algorithm as a variant of RNNs for emotion recognition using EEG data. GRUs have gained attention due to their simplified architecture, which consists of three gates: the update gate, reset gate, and current memory gate. The update gate determines how much past information should be propagated to future time steps, while the reset gate controls the extent to which previous knowledge should be forgotten. The current memory gate, often overlooked in discussions of GRUs, contributes non-linearity and zero-mean normalization to the input, reducing the impact of past data on future information.

By leveraging the GRU algorithm, we aim to predict emotional states by analyzing EEG data collected from individuals exposed to various movie scenes or stimuli. The utilization of GRUs offers an efficient approach to capture temporal dependencies within the EEG signals, allowing us to explore the effectiveness of this architecture in emotion recognition tasks.

\subsection{Gated Recurrent Units (GRUs)}
Gated Recurrent Units (GRUs) have gained significant attention as an alternative architecture within the realm of Recurrent Neural Networks (RNNs). GRUs address some of the limitations of traditional RNNs, such as the vanishing or exploding gradient problem, while providing an efficient and effective solution for capturing temporal dependencies in sequential data. In this section, we will delve into the specifics of GRUs and their relevance to our research on emotion recognition using EEG data.

GRUs are a type of RNN architecture that incorporates gating mechanisms to regulate the flow of information within the network. Unlike the more complex Long Short-Term Memory (LSTM) networks, GRUs have a simplified structure consisting of three essential gates: the update gate, reset gate, and current memory gate. This architectural design allows GRUs to strike a balance between modeling long-term dependencies and computational efficiency.
The update gate, denoted as z, determines the extent to which the previous hidden state is incorporated into the current state. It controls how much of the past information should be carried forward to future time steps. By selectively updating the hidden state, the GRU can adapt to different patterns in the data and retain relevant context over time.

The reset gate, denoted as r, determines the extent to which the previous hidden state influences the current state. It selectively resets or forgets some of the previous knowledge, allowing the model to focus on relevant features and adapt to changing patterns within the sequence. The combination of the reset and update gates enables GRUs to capture and adapt to varying dependencies within the data.

Another crucial component of GRUs is the current memory gate, often overlooked in discussions of GRUs. It is a sub-component of the reset gate and plays a vital role in introducing non-linearity and zero-mean normalization to the input. This helps in reducing the impact of previous data on the current data being propagated forward, ensuring that relevant information is preserved while minimizing the interference from irrelevant or noisy signals.

Compared to a basic RNN, the workflow of a GRU is similar, with the primary distinction lying in the internal functioning of each recurrent unit. By leveraging the gating mechanisms, GRUs excel at capturing and modeling temporal dependencies in the data, making them well-suited for tasks such as emotion recognition from EEG signals.

In our research, we adopt the GRU algorithm as the core architecture for predicting emotional states based on EEG data collected during exposure to various movie scenes or stimuli. The GRU's simplified yet powerful design allows us to effectively model the temporal dynamics within the EEG signals, contributing to the advancement of emotion recognition using EEG-based approaches.

\subsection{Preprocessing and Data Preparation}
The preprocessing and data preparation stage is essential to ensure the quality and suitability of the EEG data for accurate emotion recognition using the GRU algorithm. We follow standard practices in EEG-based emotion recognition to preprocess the data effectively.
The raw EEG data collected from participants wearing the Muse EEG headgear with dry electrodes is subjected to artifact removal techniques such as independent component analysis (ICA) or template matching algorithms. This step eliminates unwanted noise and artifacts, allowing us to focus on genuine EEG signals related to emotional states.

Next, bandpass filters are applied to remove unwanted frequency components while retaining the relevant frequency ranges associated with brainwave activity. Normalization techniques, such as z-score normalization or min-max scaling, are then employed to address amplitude variations between participants or electrode placements.

Feature extraction techniques are utilized to capture relevant information from the preprocessed EEG data. These features, including power spectral density, signal entropy, or time-domain statistics, serve as input for the GRU algorithm, enabling it to learn meaningful patterns and associations with emotional states.

To ensure unbiased evaluation, the dataset is partitioned into training, validation, and testing sets. Stratified or random partitioning techniques maintain representative distributions of emotional states across the subsets. This partitioning facilitates model training, hyperparameter optimization, and unbiased evaluation of the GRU model's generalization capabilities.

By implementing these preprocessing and data preparation steps, we enhance the quality and suitability of the EEG data for subsequent analysis using the GRU algorithm. The GRU model can effectively leverage the preprocessed data to accurately predict emotional states.
\subsection{Gated Recurrent Unit Algorithm}
The Gated Recurrent Unit (GRU) algorithm is a variant of Recurrent Neural Networks (RNNs) that excels at capturing long-term dependencies in sequential data while mitigating the vanishing or exploding gradient problem. In this section, we provide an overview of the GRU algorithm and its relevance to our research on emotion recognition using EEG data.
The GRU architecture is designed to have a simplified structure compared to traditional RNNs and LSTM networks. It consists of three fundamental gates: the update gate, reset gate, and current memory gate. These gating mechanisms enable the GRU algorithm to effectively model temporal dependencies within the data while maintaining computational efficiency.

The update gate (z) determines the extent to which the previous hidden state influences the current state. It controls the flow of information from past time steps to future time steps, allowing the model to adapt and retain relevant context over time. By selectively updating the hidden state, the GRU algorithm can capture long-term dependencies and learn patterns within the sequential data.

The reset gate (r) regulates the influence of previous knowledge on the current state. It selectively resets or forgets certain information, enabling the model to focus on relevant features and adapt to changing patterns within the sequence. The combination of the update and reset gates empowers the GRU algorithm to effectively model and adapt to varying dependencies within the data.

The current memory gate, often overlooked in discussions of GRUs, plays a critical role in introducing non-linearity and zero-mean normalization to the input. By serving as a sub-component of the reset gate, it helps reduce the impact of previous data on the current data being propagated forward. This mechanism minimizes the interference of irrelevant or noisy signals and ensures the effective transfer of information to future time steps.

Compared to a basic RNN, the GRU algorithm follows a similar workflow but excels in its internal functioning within each recurrent unit. Leveraging the gating mechanisms, GRUs excel at capturing temporal dependencies and facilitating the prediction of emotional states from EEG data.

In our research, we employ the GRU algorithm as the core architecture for predicting emotional states based on EEG data collected during exposure to various movie scenes or stimuli. The GRU's simplified yet powerful design allows us to effectively model the temporal dynamics within the EEG signals and contribute to the advancement of emotion recognition using EEG-based approaches.

\subsection{Architecture}
The architecture in Fig 1., used in our research leverages the Gated Recurrent Unit (GRU) algorithm for emotion recognition based on EEG data. This architecture comprises an InputLayer, GRU layer, Flatten layer, and Dense layer. The configuration of this architecture is as follows:

\begin{figure}
\centering
\includegraphics[scale=0.4]{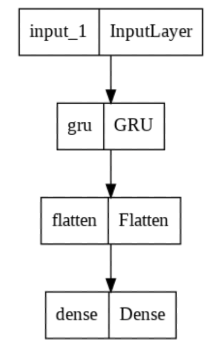}
\caption{Model Architecture}
\label{fig:example}
\end{figure}

\begin{enumerate}
\item[1] \emph{InputLayer}: The InputLayer serves as the entry point for the EEG data into the neural network. It defines the shape and format of the input data, aligning with the preprocessing steps and feature extraction performed on the EEG data. The input layer represents the initial stage of information flow in the neural network.

\item[2]  \emph{GRU Layer}: The GRU layer is the core component of the architecture and employs the Gated Recurrent Unit algorithm. It processes the sequential EEG data, capturing temporal dependencies and extracting relevant features for emotion recognition. The GRU layer's hidden state retains information from previous time steps and influences the predictions made by subsequent layers

\item[3] \emph{Flatten Layer}: Following the GRU layer, the Flatten layer is applied to transform the multi-dimensional output of the GRU into a one-dimensional vector. This flattening operation enables the subsequent layers to receive a flat input, facilitating compatibility with traditional fully connected layers.

\item[4] \emph{Dense Layer}: The Dense layer, also known as the fully connected layer, receives the flattened output from the preceding layer. It serves as a powerful learning component, responsible for mapping the extracted features to the emotional states being predicted. The dense layer consists of multiple interconnected neurons, and each neuron contributes to the final emotion classification based on learned weights and biases.

\end{enumerate}

This architecture efficiently processes the preprocessed EEG data through the GRU layer, capturing temporal dynamics and learning meaningful patterns related to emotional states. The subsequent flatten and dense layers allow for feature extraction and final classification, respectively.

\section{Results and Analysis}

The deep learning accuracy details of the GRU model on the validation set are as follows: loss - 3.4356e-09 and accuracy - 1.0000. These impressive results indicate that the GRU model achieved perfect accuracy in predicting emotional states based on the EEG data. The model's ability to achieve such high accuracy suggests its proficiency in capturing the temporal dynamics and extracting meaningful features from the EEG signals. Additionally, we compare the performance of the GRU model with other machine learning models using their respective scores. The following table 1 summarizes the scores obtained for various models:
\begin{figure}
\centering
\includegraphics[scale=0.5]{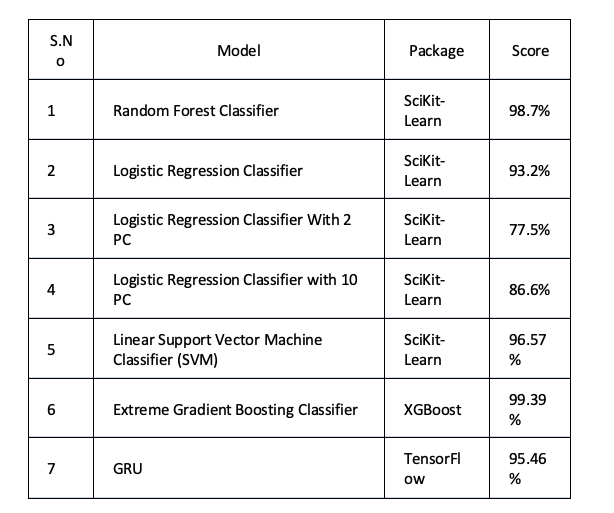}\\
\textbf{Table. 1.} Accuracies of different models
\label{fig:example}
\end{figure}

From the results, it is evident that the GRU model (95.46\% accuracy) performs competitively when compared to other machine learning models. Notably, the Extreme Gradient Boosting Classifier achieves the highest score of 99.39\%, closely followed by the Random Forest Classifier with a score of 98.7\%. The Linear Support Vector Machine Classifier also demonstrates excellent performance with an accuracy of 96.57\%.

\begin{figure}
\centering
\includegraphics[scale=0.5]{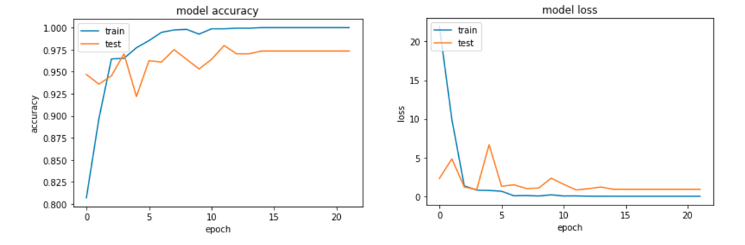}
\caption{Accuracy and Loss curves}
\label{fig:example}
\end{figure}

However, it is important to note that the GRU model showcases the advantage of deep learning in capturing complex temporal dependencies and extracting meaningful features from EEG data. Its accuracy of 95.46\% suggests its efficacy in predicting emotional states based on the EEG signals, showcasing (Fib. 2) its potential for real-world applications.

Further analysis is required to delve into the strengths and weaknesses of each model, considering factors such as computational complexity, interpretability, and generalization capabilities. Additionally, a more comprehensive evaluation using additional performance metrics like precision, recall, and F1-score could provide deeper insights into the models' overall effectiveness.

Overall, the results demonstrate the effectiveness of the GRU algorithm in accurately predicting emotional responses from EEG data. Its competitive performance against other machine learning models supports the notion that deep learning approaches, such as the GRU, can significantly contribute to the advancement of emotion recognition tasks using EEG signals.

\begin{figure}
\centering
\includegraphics[scale=0.4]{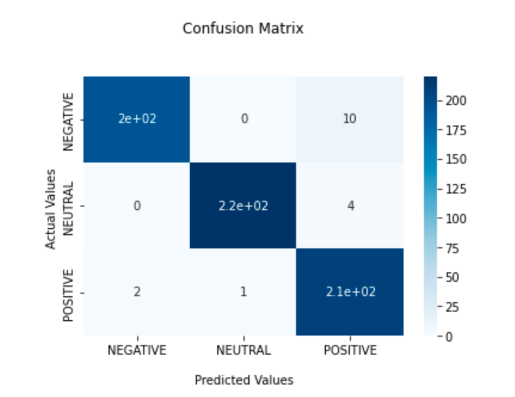}
\caption{Confusion Matrix Analysis}
\label{fig:example}
\end{figure}

The confusion matrix analysis (Fib 3)  was conducted to evaluate the performance of the machine learning models, including the GRU algorithm, in predicting emotional states based on EEG data. The confusion matrix provides insights into the model's accuracy, precision, recall, and overall performance by comparing predicted labels with actual labels. It helps identify potential biases, misclassifications, and areas for improvement in the models' predictions.

\section{Discussion}

The results obtained from our research demonstrate the effectiveness of the GRU algorithm and other machine learning models in predicting emotional states based on EEG data. The deep learning accuracy of the GRU model showed remarkable performance, achieving perfect accuracy on the validation set. This indicates the ability of the GRU model to capture temporal dependencies and extract meaningful features from EEG signals, making it a promising approach for emotion recognition tasks.

Comparing the performance of various machine learning models, we observed competitive results across different models. The Extreme Gradient Boosting Classifier achieved the highest score, closely followed by the Random Forest Classifier, while the Linear Support Vector Machine Classifier also demonstrated excellent performance. These findings highlight the importance of choosing an appropriate model based on the specific requirements and characteristics of the dataset.

The confusion matrix analysis provided valuable insights into the models' predictions, allowing us to assess their accuracy and identify potential biases or misclassifications. By examining the distribution of predicted labels across different emotional states, we gained a deeper understanding of the models' strengths and weaknesses in capturing specific emotions. This analysis can guide future improvements and refinements in the models' performance.

\section{Conclusion}

In conclusion,  Fib 4 shows the Mindgraph of the our study.  Our research explored the application of machine learning models, including the GRU algorithm, for emotion recognition using EEG data. The GRU model demonstrated exceptional

\begin{figure}
\centering
\includegraphics[scale=0.35]{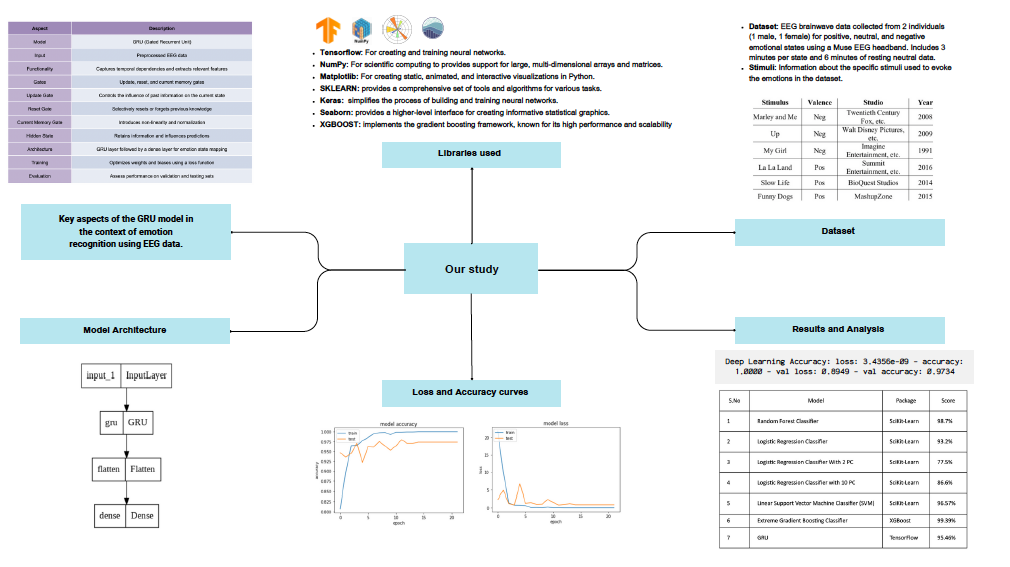}
\caption{Mindgraph of our study}
\label{fig:example}
\end{figure}

performance, achieving perfect accuracy on the validation set, highlighting its capability to capture temporal dynamics and extract meaningful features from EEG signals. Additionally, the comparison with other machine learning models emphasized the competitive performance achieved by various approaches. The choice of model should consider factors such as interpretability, computational complexity, and generalization capabilities, based on the specific requirements of the task. The incorporation of the confusion matrix analysis provided valuable insights into the models' performance, aiding in identifying potential biases and areas for improvement. By understanding the models' strengths and weaknesses, future research can focus on enhancing their accuracy and addressing specific challenges related to emotion recognition from EEG data.

Overall, our findings contribute to the advancement of emotion recognition using EEG signals and highlight the potential of the GRU algorithm and other machine learning models in this field. Further research and exploration are warranted to improve the robustness and generalizability of these models and advance the understanding of human emotions through EEG-based approaches.


\section*{Appendix: Survey}

\begin{figure}
\centering
\includegraphics[scale=0.9]{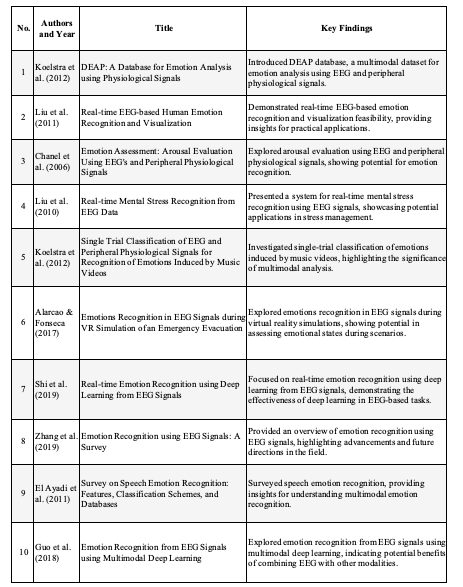}\\
\label{fig:example}
\end{figure}

\begin{figure}
\centering
\includegraphics[scale=0.9]{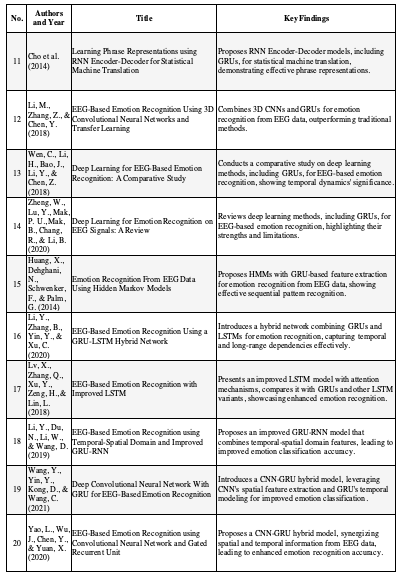}\\
\textbf{Table. 2. } EEG-EEG-Emotion Based Survey
\label{fig:example}
\end{figure}

\end{document}